\documentclass[10pt,a4,conference]{IEEEtran}
\usepackage{cite}
\usepackage{subfig}
\usepackage{balance}
\usepackage{tabularx}
\usepackage{booktabs}
\usepackage{amsmath,amssymb,amsfonts}
\usepackage{mathtools} 
\usepackage{graphicx}
\usepackage{xcolor}
\usepackage{subfig}
\usepackage{tikz}
\usepackage{float}
\usepackage{subfig}
\usepackage[nolist,nohyperlinks]{acronym}
\usepackage{algorithm}
\usepackage{algpseudocode}
 \usepackage{setspace}
\let\Algorithm\algorithm
\renewcommand\algorithm[1][]{\Algorithm[#1]\setstretch{1.1}}

\newcommand{\av}{\mathbf{a}}

\newcommand{\thetar}{\theta_{\mathrm{R}}}
\newcommand{\thetarn}{\theta^{(n)}_{\mathrm{R}}}
\newcommand{\thetan}{\vartheta_{n}}
\newcommand{\Thetaone}{\boldsymbol{\Theta}_1}
\newcommand{\Thetatwo}{\boldsymbol{\Theta}_2}
\newcommand{\Na}{N_{\mathrm{a}}}
\newcommand{\Nt}{N_{\mathrm{T}}}
\newcommand{\Nr}{N_{\mathrm{R}}}
\newcommand{\Nbs}{N_{\mathrm{BS}}}
\newcommand{\h}{\mathbf{h}}
\newcommand{\rv}{\mathbf{r}}
\newcommand{\hchan}{\mathbf{H}}

\newcommand{\y}{\mathbf{y}}

\newcommand{\fd}{f_\mathrm{D}}
\newcommand{\Ts}{T_\mathrm{s}}
\newcommand{\Df}{\Delta f}
\newcommand{\xrn}{x^{(n)}_{\mathrm{R}}}
\newcommand{\yrn}{y^{(n)}_{\mathrm{R}}}

\newcommand{\EX}[1] {{\mathbb{E}}\left\{{#1}\right\}}
\newcommand{\p}{\mathbf{p}}
\newcommand{\ph}{\widehat{\mathbf{p}}}
\newcommand{\I}{\boldsymbol{\mathcal{I}}}
\newcommand{\Ie}{\boldsymbol{\mathcal{I}}_\mathrm{e}}
\newcommand{\Jm}{\mathbf{J}_\mathrm{m}}

\newcommand{\Jn}{\mathbf{J}_\mathrm{n}}

\newcommand{\SNR}{\mathrm{SNR}}
\newcommand{\CRB}{\mathrm{CRB}}
\newcommand{\herm}{\mathsf{H}}
\newcommand{\transp}{\mathsf{T}}

\newcounter{MYtempeqncnt}

\DeclareMathOperator{\diag}{diag}
\DeclareMathOperator{\Tr}{Tr}

\begin{acronym}
\acro{AWGN}{additive white Gaussian noise}
\acro{AoA}{angle of arrival}
\acro{BF}{beamformer}
\acro{BS}{base station}
\acro{CP}{cyclic prefix}
\acro{CRLB}{Cram\'{e}r-Rao lower bound}
\acro{DoA}{direction of arrival}
\acro{DoD}{direction of departure}
\acro{EIRP}{effective isotropic radiated power}
\acro{ELP}{equivalent low-pass}
\acro{EFIM}{equivalent Fisher information matrix}
\acro{FFT}{fast Fourier transform}
\acro{FIM}{Fisher information matrix}
\acro{IFFT}{inverse fast Fourier transform}
\acro{i.i.d.}{independent, identically distributed}
\acro{JSC}{joint sensing and communication}
\acro{LoS}{line-of-sight}
\acro{LTE}{long term evolution}
\acro{MCL}{maximum coupling loss}
\acro{MPL}{maximum path loss}
\acro{MIL}{maximum isotropic loss}
\acro{MC}{Monte Carlo}
\acro{MIMO}{multiple-input multiple-output}
\acro{MUSIC}{MUltiple SIgnal Classification}
\acro{MDL}{minimum description length}
\acro{NR}{new radio}
\acro{OFDM}{orthogonal frequency division multiplexing}
\acro{OSPA}{optimal sub-pattern assignment}
\acro{PEB}{position error bound}
\acro{PSD}{power spectral density}
\acro{QPSK}{quadrature phase shift keying}
\acro{RCS}{radar cross-section}
\acro{RF}{radio frequency}
\acro{RMSE}{root mean square error}
\acro{r.v.}{random variable}
\acro{Rx}{receiver}
\acro{SSIR}{signal-to-self interference ratio}
\acro{SI}{self interference}
\acro{SNR}{signal-to-noise ratio}
\acro{SU}{single-user}
\acro{SCM}{sample covariance matrix}
\acro{Tx}{transmitter}
\acro{UE}{user equipment}
\acro{ULA}{uniform linear array}
\end{acronym}

\IEEEoverridecommandlockouts 
\begin{document}

\title{Position Error Bound for Cooperative Sensing in MIMO-OFDM Networks
\thanks{This work was supported by the European Union under the Italian National Recovery and Resilience Plan (NRRP) of NextGenerationEU, partnership on ``Telecommunications of the Future'' (PE00000001 - program ``RESTART''). The authors are with the Wireless Communications Laboratory, CNIT, DEI, University of Bologna, Italy,  
Email: \{lorenzo.pucci3, andrea.giorgetti\}@unibo.it
}
}
\author{
\IEEEauthorblockN{Lorenzo Pucci and Andrea~Giorgetti}
}
\maketitle

\begin{abstract}
This paper investigates the fundamental limits of target position estimation accuracy of \ac{JSC} networks comprising several monostatic \acp{BS} that cooperate to localize targets. Specifically, each \ac{BS} adopts a \ac{MIMO}-\ac{OFDM} scheme with a multi-beam radiation pattern to partition power between communication and sensing tasks. Building on prior works, we derive a general framework to evaluate the positioning accuracy of a target in networks with an arbitrary number of cooperating \acp{BS} and arbitrary geometrical configurations using Fisher information. Numerical results demonstrate the benefits of cooperation between \acp{BS} in improving target localization accuracy and provide insights into the relationships between various system parameters, which may aid in designing \ac{JSC} networks.
\end{abstract}
\acresetall


\section{Introduction}
\Ac{JSC}, namely the ability of mobile communication systems to jointly communicate and sense the environment, has emerged as a new paradigm for the upcoming 6G mobile networks. Research has recently demonstrated the effectiveness of \ac{OFDM}-based communication systems in performing sensing tasks. In particular, several works have investigated the sensing performance of \ac{OFDM}-based \ac{JSC} systems in terms of \ac{RMSE} of target parameters estimation, mainly with monostatic and bistatic configurations \cite{Braun, TagMizetal:J23, PucMatPaoGio:C22}.
Moreover, several works have investigated the performance of \ac{JSC} systems in terms of \ac{CRLB}, mainly considering monostatic radar settings. In \cite{liu2021cramer} and \cite{ZabPaoXuGio:C22} narrowband monostatic \ac{JSC} systems are considered. The \ac{CRLB} is used as a performance metric of target estimation to study the trade-off between communication and sensing. In \cite{Giovannetti2024PEB}, the \ac{CRLB} of position estimation is derived considering a monostatic \ac{JSC} system in a vehicular scenario. However, to the best of the authors' knowledge, there remains a gap in research regarding the investigation of fundamental limits of target position estimation accuracy by considering \ac{JSC} networks in which multiple monostatic sensors cooperate to improve estimation accuracy.

Aiming to fill this gap, this work investigates the fundamental limits of target position estimation accuracy in a \ac{OFDM}-based \ac{JSC} network comprising monostatic \ac{MIMO} \acp{BS}. These \acp{BS} cooperate to localize targets within the monitored area, using a multi-beam radiation pattern to share the transmit power between communication and sensing functions. Inspired from previous works \cite{shen2010fundamental1,shen2010fundamental,SheDaiWin:J14}, we derive a framework to calculate the \ac{PEB} of target localization accuracy via Fisher information.
In particular, we start deriving a closed-form solution to assess the \ac{CRLB} of target position accuracy when only one \ac{BS} is involved, using the concept of \ac{EFIM}. Then, we develop a general formulation for the \ac{PEB} with multiple cooperative \acp{BS} by exploiting the additivity of Fisher information. Based on the theoretical framework, we examine the sensing coverage of the \ac{JSC} network and analyze factors affecting localization accuracy, including the number of \acp{BS}, available bandwidth, number of antennas, and power allocation. 

In this paper, we use capital boldface letters for matrices and lowercase bold letters for vectors. Additionally, $(\cdot)^\transp$ and $(\cdot)^\herm$ denote transpose and conjugate transpose, respectively; $\bigl\| \cdot \bigr\|$ is the Euclidean norm of vectors; $[\mathbf{X}]_{a:b,c:d}$ denotes a submatrix of $\mathbf{X}$ composed of rows $a$ to $b$ and columns $c$ to $d$; $\mathbf{I}_n$ is the $n \times n$ identity matrix, while $\Tr(\cdot)$ is the trace of a square matrix. $\mathbb{E}\{\cdot\}$ and $\mathbb{V}(\cdot)$ represent mean value and variance, respectively. A zero-mean circularly symmetric complex Gaussian random vector with covariance $\boldsymbol{\Sigma}$ is denoted by $\mathbf{x} \thicksim \mathcal{CN}( \mathbf{0},\boldsymbol{\Sigma})$. $| \cdot |$ is the absolute value operator.

The paper is organized as follows. Section~\ref{sec:intro} and Section~\ref{sec:PEB} present the system model and the \ac{PEB} derivation, respectively. Numerical results are given in Section~\ref{sec:numres}, while Section~\ref{sec:conclu} concludes the paper with some remarks.

\section{System Model for Monostatic JSC}\label{sec:intro}
In this analysis, a network of monostatic \ac{OFDM}-based \ac{JSC} systems with multiple antennas is considered. Each \ac{BS} consists of a \ac{Tx} antenna array with $N_\mathrm{T}$ elements and of an \ac{Rx} antenna array with $\Nr$ elements. 
The system transmits a waveform with $M$ \ac{OFDM} symbols and $K$ active subcarriers. The \ac{ELP} representation of the signal transmitted by the $n$th antenna can be written as
\begin{equation}
       \label{eq:base-band}
        s_n(t) = \sum_{m=0}^{M-1}\left( \sum_{k=0}^{K-1}x_n[k,m]e^{j2 \pi \frac{k}{T}t}\right)g(t-m\Ts)  
\end{equation}
where $x_n[k,m]$ is the complex modulation symbol to be transmitted on the $k\text{th}$ subcarrier at time instant $m$, mapped through digital precoding at the $n\text{th}$ transmitting antenna. Moreover, $g(t)$ is the employed pulse, $\Df = 1/T$ is the subcarrier spacing, and $\Ts=T+T_\mathrm{CP}$ is the \ac{OFDM} symbol duration including the \ac{CP}. Without loss of generality, the constellation is normalized so that $\EX{|x_n[k,m]|^2}=1$.  

\subsection{Dual-functional waveform}\label{sec:waveform}
We consider the case where the same symbol $x_n[k,m]$ is transmitted to all antennas via beamforming, i.e., $x_n[k,m]=x[k,m]$. Therefore, the vector collecting the signal samples at the antennas is $\mathbf{x}[k,m] = \mathbf{w}_\mathrm{T} x[k,m] \in \mathbb{C}^{N_\mathrm{T} \times 1}$, where $\mathbf{w}_\mathrm{T} \in \mathbb{C}^{N_\mathrm{T} \times 1}$ is the precoding vector, or \ac{BF}.

In this work, the vector $\mathbf{w}_\mathrm{T}$ is designed to have a multibeam radiation pattern so that the total transmit power of the \ac{OFDM} signal is split between communication and sensing beams \cite{zhang2018multibeam, barneto2020multibeam}. Thus, the transmitting \ac{BF} $\mathbf{w}_\mathrm{T}$ can be written as
\begin{equation} 
\mathbf{w}_{\mathrm{T}} = \sqrt{\rho}\,\mathbf{w}_\mathrm{T,s} + \sqrt{1-\rho}\,\mathbf{w}_\mathrm{T,c}
\label{eq:BFvector}
\end{equation}
where $\rho \in [0,1]$ is the parameter used to control the fraction of the total power apportioned to the two directions, while $\mathbf{w}_\mathrm{T,c}$ and $\mathbf{w}_\mathrm{T,s}$ are the communication and the sensing \acp{BF}, respectively. Without loss of generality, we employ beam-steering for both sensing and communication tasks. The literature also presents other beamforming methods, such as those based on optimization techniques, which can enhance performance further \cite{Luoetal:J19}. Accordingly, the \acp{BF} are defined as \cite{asplund2020advanced} 
\begin{equation}
\mathbf{w}_\mathrm{T,c} = \frac{\sqrt{P_\mathrm{avg} G_\mathrm{T}^\mathrm{a}}}{N_\mathrm{T}}\,\mathbf{a}(\theta_\mathrm{T,c}), \;
\mathbf{w}_\mathrm{T,s} = \frac{\sqrt{P_\mathrm{avg}  G_\mathrm{T}^\mathrm{a}}}{N_\mathrm{T}}\,\mathbf{a}(\theta_\mathrm{T,s})
\label{eq:txBF}
\end{equation}
where $G_\mathrm{T}^\mathrm{a}$ is the transmit array gain along the beam steering direction (where such a gain is maximum), $P_\mathrm{avg} G_\mathrm{T}^\mathrm{a}$ is the average \ac{EIRP} per subcarrier, being $P_\mathrm{avg} = P_\mathrm{T}/K$ with $P_\mathrm{T}$ the total transmit power. Moreover, $\mathbf{a}(\theta_\mathrm{T,c}) \in \mathbb{C}^{N_\mathrm{T} \times 1}$ and $\mathbf{a}(\theta_\mathrm{T,s}) \in \mathbb{C}^{N_\mathrm{T} \times 1}$ are the steering vectors related to communication and sensing, respectively. Considering \ac{ULA} with $N_\mathrm{a}$ elements half-wavelength spaced and taking its center as a reference, the spatial steering vector at a given \ac{DoA}/\ac{DoD} $\theta$ is \cite[Chapter~9]{Richards}
\begin{equation}
\av(\theta) = \big[e^{-\imath \frac{\Na-1}{2}\pi \sin \theta},\dots,e^{\imath \frac{\Na-1}{2}\pi \sin \theta}\big]^\transp. 
\end{equation}
%
%
\begin{figure}[t]
\captionsetup{font=footnotesize,labelfont=footnotesize}
    \centering
    \includegraphics[width=0.75\columnwidth]{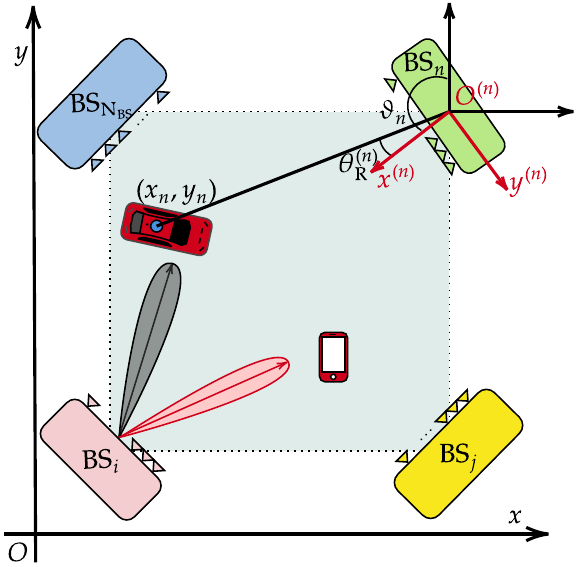}
    \caption{An example of a cooperative \ac{JSC} network of interest.}
    \label{fig:cooperative_scenario}
\end{figure}

\subsection{Received signal} \label{sec:rxSignal}
The vector $\rv[k,m] \in \mathbb{C}^{N_\mathrm{R} \times 1}$ of the received symbols after the \ac{FFT} block in the \ac{OFDM} receiver, is given by\footnote{This notation represents a 3D array where each vector contains samples along the antennas, with one vector corresponding to each $(k,m)$ pair.}
\begin{equation}
    \mathbf{r}[k,m] = \hchan[k,m]\mathbf{x}[k,m] + \boldsymbol{\nu}[k,m] 
    \label{eq:y}
\end{equation}
where $\hchan[k,m] \in \mathbb{C}^{\Nr \times \Nt}$ is the \ac{MIMO} channel matrix for the $k$th subcarrier at time $m$ and $\boldsymbol{\nu}[k,m]\sim \mathcal{CN}(\mathbf{0},\sigma_\mathrm{N}^2 \mathbf{I}_{\Nr})$ is the \ac{AWGN} at \ac{Rx} antennas with $\sigma_\mathrm{N}^2 = N_0 \Delta f$. Here $N_0 = k_\mathrm{B} T_0 n_\mathrm{F}$ is the one-sided noise \ac{PSD},  being $k_\mathrm{B}$ the Boltzmann constant, $T_0$ the reference temperature and $n_\mathrm{F}$ the receiver noise figure. 

Considering a tapped delay line channel model with reflections from $L$ scatterers and \ac{LoS} propagation conditions, the channel matrix can be written as
\begin{equation}\label{eq:channel-matrixFull}
    \hchan[k,m] = \sum_{l = 1}^{L} \underbrace{\bar{\alpha}_l e^{\imath 2\pi m \Ts f_{\mathrm{D},l}}e^{-\imath 2\pi k \Df \tau_l}}_{\triangleq \beta_l} \mathbf{b}(\theta_{\mathrm{R},l})\mathbf{a}^\herm(\theta_{\mathrm{R},l})
\end{equation}
where $\tau_l$, $f_{\mathrm{D},l}$, and $\theta_{\mathrm{R},l}$ are the round-trip delay, the Doppler shift, and the \ac{DoA} of the $l$th target, respectively. The term $\bar{\alpha}_l = \alpha_l e^{j\phi_l}$ is the complex amplitude which includes phase shift and attenuation along the $l$th propagation path. According to the radar equation, $\alpha^2_l~= \frac{G_\mathrm{R} c^2 \sigma_l }{(4 \pi)^3 f^2_\mathrm{c}\, r_l^4}$, being $c$ the speed of light, $f_\mathrm{c}$ the carrier frequency, $r_l$ and $\sigma_l$ the distance between the monostatic system and the scatterer $l$ and the radar cross-section of the $l$th scatterer, respectively; $G_\mathrm{R}$ is the gain of the single \ac{Rx} antenna element.
Considering a single scatterer scenario, i.e., $L=1$, and dropping the index $l$, the received signal in \eqref{eq:y} simplifies into
%
\begin{equation}
    \mathbf{r}[k,m] = \beta \mathbf{b}(\theta_{\mathrm{R}})\mathbf{a}^\herm(\theta_{\mathrm{R}})\mathbf{w}_\mathrm{T} x[k,m] + \boldsymbol{\nu}[k,m]. 
    \label{eq:y2}
\end{equation}

After removing the transmitted symbols, $x[k,m]$ (assumed to be known), which represent nuisance, through element-wise division, the received samples are obtained\footnote{This approach is often called reciprocal filtering.}
\begin{equation}\label{eq:recsigdiv}
     \y[k,m]= \beta \h + \mathbf{n}[k,m]
\end{equation}
where $\mathbf{n}[k,m]\sim 
\mathcal{CN}(\mathbf{0},\widetilde{\sigma}_\mathrm{N}^2[k,m] \mathbf{I}_{\Nr})$ with $\widetilde{\sigma}_\mathrm{N}^2[k,m]=\sigma_\mathrm{N}^2/|x[k,m]|^2$, and $\h \triangleq \mathbf{b}(\theta_{\mathrm{R}}) \mathbf{a}^\herm(\theta_{\mathrm{R}}) \mathbf{w}_\mathrm{T}$.\footnote{Note that if a constant envelope modulation is used, e.g., \ac{QPSK} then $\widetilde{\sigma}_\mathrm{N}^2[k,m]=\sigma_\mathrm{N}^2$.} Recalling that the noise after \ac{FFT} blocks is white across the time, frequency, and spatial (antenna) domain, the log-likelihood function of the received ensemble  $\boldsymbol{\mathcal Y}=\{\y[k,m]\}_{k=0,\dots,K-1,m=0,\dots,M-1}$ is
%
%
\begin{figure*}[t]
\setcounter{MYtempeqncnt}{\value{equation}}
\setlength{\arraycolsep}{2pt}
\begin{equation}\label{eq:FIMofdm}
\I(\boldsymbol{\Theta})=\Gamma
\times\begin{bmatrix}
\frac{2}{\alpha^2} & 0 & 0 & 0 & 0\\
0 & 2 & 2\pi  \Ts (M-1)
 & -2\pi \Df (K-1)
& 0\\
0 & 2\pi \Ts (M-1)
 & \frac{4 \pi^2 \Ts^2 (2M-1)(M-1)}{3} & -2\pi^2 \Ts \Df (M-1)(K-1) & 0\\
0 & -2\pi \Df (K-1) & -2\pi^2 \Ts \Df (M-1)(K-1) & \frac{4\pi^2 \Df^2 (2K-1)(K-1)}{3} & 0\\
0 & 0 & 0 & 0 & \frac{\pi^2 (\Nr^2-1)\cos^2(\thetar)}{6}
\end{bmatrix}
\end{equation}
\hrulefill
\end{figure*}

\begin{align}\label{eq:LLF}
\ln f(\boldsymbol{\mathcal Y})  
& = -\sum_{k=0}^{K-1} \sum_{m=0}^{M-1} \Nr \ln \widetilde{\sigma}_\mathrm{N}^2[k,m] \\
&\qquad \qquad \qquad +\frac{1}{\widetilde{\sigma}_\mathrm{N}^2[k,m]}\Bigl\|\y[k,m]\Bigr.\Bigl.-\beta  \h\Bigr\|^2
\nonumber
\end{align}
where we omitted those terms that are not relevant for the derivation of the \ac{FIM}.

\section{Position Error Bound With and Without Cooperation}\label{sec:PEB}

\subsection{Fisher information matrix and CRLB}\label{sec:CRLB}
The \ac{FIM} of the vector of parameters ${\boldsymbol{\Theta}}=(\alpha,\phi,\fd,\tau,\theta_\mathrm{R})^\transp$ is calculated from \eqref{eq:LLF} as
\begin{align}\label{eq:FIMdef}
&[\I(\boldsymbol{\Theta})]_{i,j}=-\EX{\frac{\partial^2 \ln f(\boldsymbol{\mathcal Y})}{\partial \theta_i \partial\theta_j}} \quad i,j=1,\dots,5 \\
& = \sum_{k=0}^{K-1} \sum_{m=0}^{M-1}\mathbb{E} \left\{ \frac{|x[k,m]|^2}{\sigma_\mathrm{N}^2} 
\frac{\partial^2}{\partial \theta_i \partial\theta_j}
\left\| \y[k,m] -\beta\h
\right\|^2\right\} \nonumber
\end{align}

By substituting \eqref{eq:recsigdiv} into \eqref{eq:FIMdef} and following lengthy calculations, the \ac{FIM} can be derived, as reported at the top of the next page in \eqref{eq:FIMofdm}, where $\Gamma = \SNR \Nr K M$. Subsequently, inverting \eqref{eq:FIMofdm} yields the expression for the \ac{CRLB} of each parameter:

\begin{align}\label{eq:IFIM}
  \CRB(\alpha)&=[\I^{-1}]_{1,1}=
  \frac{\alpha ^2}{2 K M \Nr \SNR} 
\\
  \CRB(\phi)&=[\I^{-1}]_{2,2}=
  \frac{7 K M+K+M-5}{2 (K^2+K) \left(M^2+M\right)
   \Nr \SNR}
\nonumber\\
\CRB(\fd)&=[\I^{-1}]_{3,3}=\frac{3}{2 \pi^2 \Ts^2 K M (M^2-1) \Nr
   \SNR}
\nonumber\\
  \CRB(\tau)&=[\I^{-1}]_{4,4}=
  \frac{3}{2 \pi ^2 \Delta f^2 M K (K^2-1)\Nr \SNR} 
\nonumber\\
  \CRB(\theta_\mathrm{R})&=[\I^{-1}]_{5,5}=
  \frac{6}{\pi^2 K M (\Nr^2-1) \Nr \SNR\cos^2(\thetar)}
  \nonumber.
\end{align}

By choosing $\mathbf{w}_\mathrm{T}$ according to \eqref{eq:BFvector}, setting $\theta_\mathrm{T,s} = \theta_\mathrm{R}$ and considering sufficiently large $\Nt$, the \ac{SNR} at the single receiving antenna element is given by
\begin{equation}\label{eq:SNR3}  
\SNR = \rho\frac{P_\mathrm{T}G_\mathrm{T}^a \alpha^2 }{N_0 K \Df} = \rho \frac{P_\mathrm{T}G_\mathrm{T}^a G_\mathrm{R} c^2 \sigma }{(4 \pi)^3 f^2_\mathrm{c}\, r^4 N_0 K \Df}.
\end{equation}


\vspace{-0.075cm}
\subsection{PEB for a monostatic sensor} \label{sec:monoCRLB}
Let us define the target position as $\p=(x,y)=(r\cos(\thetar),r\sin(\thetar))$, the estimated position as $\ph=(\widehat{x},\widehat{y})=(\widehat{r}\cos(\widehat{\theta}_\mathrm{R}),\widehat{r}\sin(\widehat{\theta_\mathrm{R}}))$ and the error as $e_\mathrm{p}=\|\ph-\p\|=\sqrt{(x-\widehat{x})^2+(y-\widehat{y})^2}$. The \ac{CRLB} for the target position estimation in monostatic sensing can be calculated retaining only the information of the \ac{FIM} \eqref{eq:FIMofdm} related to $\tau$ and $\thetar$ via the \ac{EFIM} \cite{shen2010fundamental}. Given the parameter vector ${\boldsymbol{\Theta}}=(\alpha,\phi,\fd,\tau,\theta_\mathrm{R})^\transp$ we partition it into ${\boldsymbol{\Theta}}=[\Thetaone^\transp,\Thetatwo^\transp]^\transp$ with $\Thetaone=(\alpha,\phi,\fd)^\transp$ $\Thetatwo=(\tau,\theta_\mathrm{R})^\transp$,
which induces the partition of the \ac{FIM} into the matrices $\mathbf{A}\in \mathbb{R}^{3\times 3}$, $\mathbf{B}\in \mathbb{R}^{3\times 2}$, and $\mathbf{C}\in \mathbb{R}^{2\times 2}$, as
%
$\I(\boldsymbol{\Theta})=
\big[\begin{smallmatrix}
\mathbf{A} & \mathbf{B} \\
\mathbf{B}^\transp & \mathbf{C}
\end{smallmatrix}\big].$
%
Then, the \ac{EFIM} is
\begin{align}\label{eq:EFIMdef}
\Ie(&\boldsymbol{\Thetatwo}) = \mathbf{C}-\mathbf{B}^\transp\mathbf{A}^{-1}\mathbf{B}  \\
& = \SNR\,\Nr K M \times
\begin{bmatrix}
\frac{2\pi^2\Df^2 (K^2-1)}{3} & 0 \\
0 & \frac{\pi^2(\Nr^2-1)cos^2(\thetar)}{6}
\end{bmatrix} \nonumber
\end{align}
known as the Schur complement of $\mathbf{A}$.\footnote{One interesting property is that $[\I^{-1}(\boldsymbol{\Theta})]_{4:5,4:5}=\Ie^{-1}(\boldsymbol{\Thetatwo}),$ so the \ac{EFIM} retains all the necessary to derive the information inequality for $\Thetatwo$.}
Since we are interested in the \ac{CRLB} of position estimation, we need a reparameterization to calculate the \ac{EFIM} in \eqref{eq:EFIMdef} for $\p$, i.e.,
%
%
\begin{equation}\label{eq:EFIMp}
\Ie(\p)=\Jm^\transp \Ie(\Thetatwo)\Jm
\end{equation}
where $\Jm$ is the Jacobian of the transformation $\{\thetar=\arctan(y/x),\tau=\frac{2}{c}\sqrt{x^2+y^2}\}$, which corresponds to 
\begin{equation}
\Jm=
\begin{bmatrix}
\frac{\partial \tau}{\partial x} & \frac{\partial \tau}{\partial y} \\
\frac{\partial \thetar}{\partial x} & \frac{\partial \thetar}{\partial y}
\end{bmatrix}
=
\begin{bmatrix}
\frac{2}{c}\frac{x}{\sqrt{x^2+y^2}} & \frac{2}{c}\frac{y}{\sqrt{x^2+y^2}} \\
-\frac{y}{x^2+y^2} & \frac{x}{x^2+y^2}
\end{bmatrix}.
\end{equation}
Hence, the \ac{CRLB} is 
%
%
%
\begin{align}\label{eq:CRLB}
\mathbb{V}\{e_\mathrm{p}\}\geq&\CRB(\p)=\Tr(\Ie^{-1}(\p)) \\
&=\Tr(\Jm^{-1}\Ie^{-1}(\Thetatwo)[\Jm^{-1}]^\transp)\nonumber\\
&=\Tr(\Ie^{-1}(\Thetatwo)[\Jm^{-1}]^\transp\Jm^{-1})=\Tr(\Ie^{-1}(\Thetatwo)\mathbf{M}_\mathrm{m}) \nonumber
\end{align}
where $\mathbf{M}_\mathrm{m}=\diag(c^2/4,x^2+y^2)$. Therefore, we get
\begin{align} \label{eq:CRB_p}
& \CRB(\p)=\frac{c^2}{4}[\I^{-1}]_{4,4}+r^2[\I^{-1}]_{5,5}\\
  &\, =\frac{6}{\pi^2 K M \Nr \SNR}\left[\frac{c^2/16}{\Df^2 (K^2-1)}+\frac{r^2}{(\Nr^2-1) \cos^2(\thetar)}\right].\nonumber
\end{align}
From \eqref{eq:CRB_p} we have
$\mathrm{PEB} = \sqrt{\CRB(\p)}$.
\subsection{PEB for a network of cooperative monostatic sensors}\label{sec:netmonoCRLB}
Let us consider a \ac{JSC} network consisting of $\Nbs$ monostatic \acp{BS} that cooperate to monitor a given area, as shown in Fig.~\ref{fig:cooperative_scenario}.\footnote{We assume that \acp{BS} can operate using either time or frequency division among themselves to avoid interference.} Let $\mathbf{O}^{(n)}=(\xrn,\yrn)$ and $\p = (x,y)$ be the position of the $n$th monostatic \ac{BS} and of a generic point-like target, respectively, in a common Cartesian reference system. Moreover, let $\p^{(n)}=(x_n,y_n)$ and $\thetarn$ be the position and the \ac{DoA} of the target, respectively, with respect to the $n$th local reference system centered at $\mathbf{O}^{(n)}$. The relation between $\p$ and $\p^{(n)}$ is $\{x_n = x'\cos(\thetan) + y'\sin(\thetan),\, y_n = -x'\sin(\thetan) + y'\cos(\thetan) \}$,
where $x'=x-\xrn$ and $y'=y-\yrn$, while $\thetan$ is the counterclockwise rotation angle of the $n$th reference system with respect to the standard one.

Considering that each of the $\Nbs$ \acp{BS} performs target position estimation independently from the others, the \ac{FIM} for cooperative position estimation is given by
\begin{equation}\label{eq:PEB_n}
    \Ie(\p) = \sum_{n=1}^{N_\mathrm{BS}} \Jn^\mathsf{T} \Ie(\p^{(n)}) \Jn
\end{equation}
where $\Ie(\p^{(n)})$ represents the \ac{EFIM} for the target position $\p^{(n)}$ as seen by the $n$th \ac{BS}. The elements of $\Ie(\p^{(n)})$ are computed according to \eqref{eq:EFIMp} by setting $(x,y)=(x_n,y_n)$ and $\thetar = \thetarn$, and given by
\begin{equation}\label{eq:PEB_n_elem}
\Ie(\p^{(n)})= \xi_n 
\begin{bmatrix}
a\, x_n^2\, r_n^2+ b \,y_n^2 & x_n \, y_n\, (a\,r_n^2-b) \\
 x_n\, y_n\,(a\, r_n^2-b) & a\, y_n^2 \, r_n^2+ b\,x_n^2
\end{bmatrix}
\end{equation}
where $a = 16 \Df^2 (K^2-1)$, $b =  c^2 (\Nr^2-1)\cos^2(\thetarn)$, $r_n$ is the Euclidean distance between the target and the $n$th \ac{BS}, and $\xi_n= \frac{\pi^2 K M \Nr \SNR_n}{6 c^2 r_n^4}$. In $\xi_n$, $\SNR_n$ is the \ac{SNR} computed according to \eqref{eq:SNR3} with respect to the $n$th \ac{BS}. Note that, $\Jn$ in \eqref{eq:PEB_n} represents the Jacobian of the transformation $n:~(x,y) ~\rightarrow~(x_n,y_n)$, defined as
\begin{equation}
\Jn=
\begin{bmatrix}
\frac{\partial x_n}{\partial x} & \frac{\partial x_n}{\partial y} \\
\frac{\partial y_n}{\partial x} & \frac{\partial y_n}{\partial y}
\end{bmatrix}
=
\begin{bmatrix}
\cos(\thetan) & \sin(\thetan)\\
-\sin(\thetan) & \cos(\thetan)
\end{bmatrix}.
\end{equation}

After computing \eqref{eq:PEB_n}, the \ac{PEB} for a system composed of $\Nbs$ \acp{BS}, which cooperate to perform target localization in a given area can be obtained as the square root of $\CRB(\p)$, computed according to the first line of \eqref{eq:CRLB}. 

\begin{figure*}[ht]
    \captionsetup{font=footnotesize,labelfont=footnotesize}
    \centering
    \subfloat[][\emph{$\Nbs = 2$} \label{fig:heatmap2}] 
    {\includegraphics[width=0.325\textwidth, keepaspectratio]{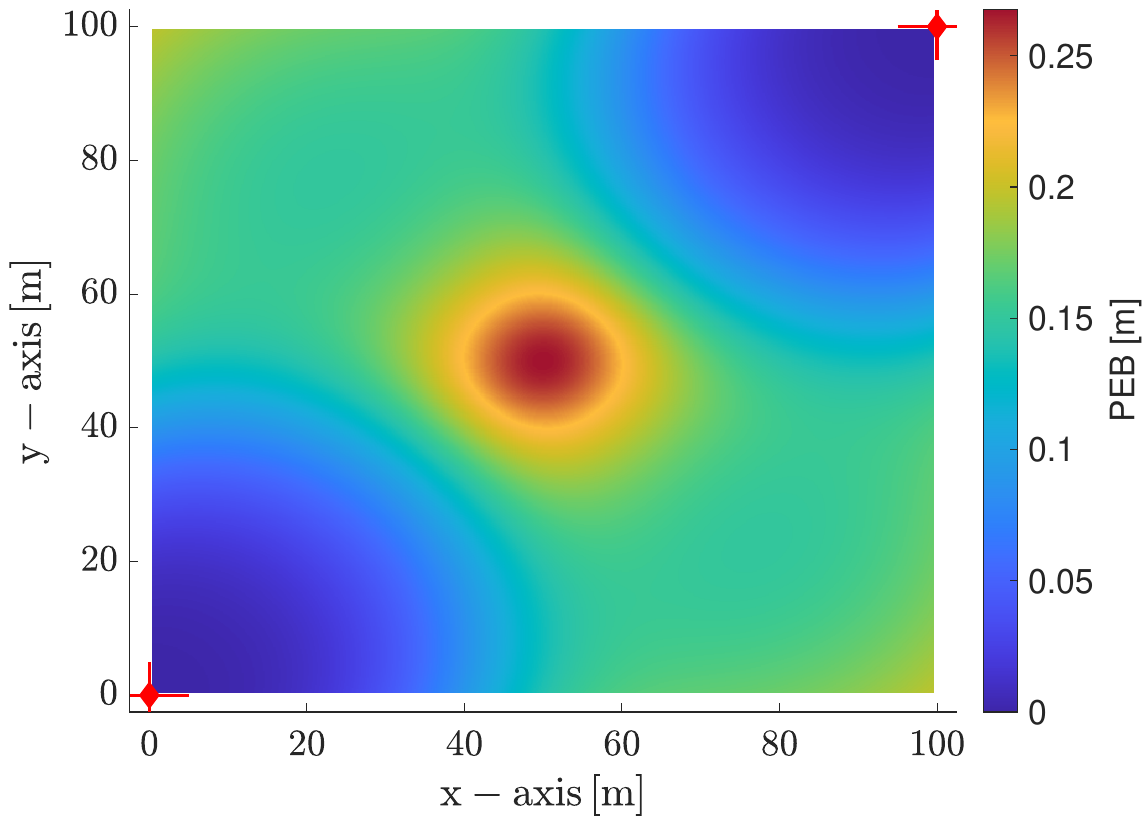}} 
    \,
    \subfloat[][\emph{$\Nbs = 3$} \label{fig:heatmap3}]
    {\includegraphics[width=0.325\textwidth, keepaspectratio]{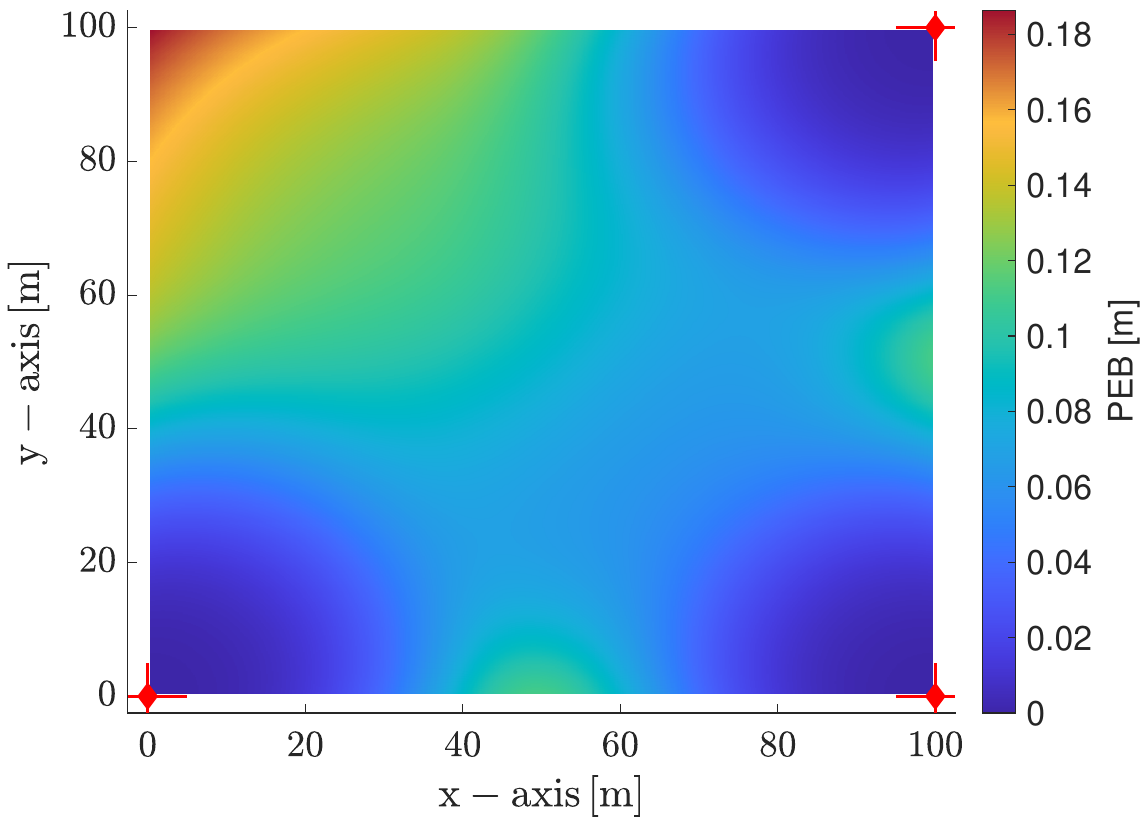}}
    \,
    \subfloat[][\emph{$\Nbs = 4$} \label{fig:heatmap4}]
    {\includegraphics[width=0.325\textwidth, keepaspectratio]{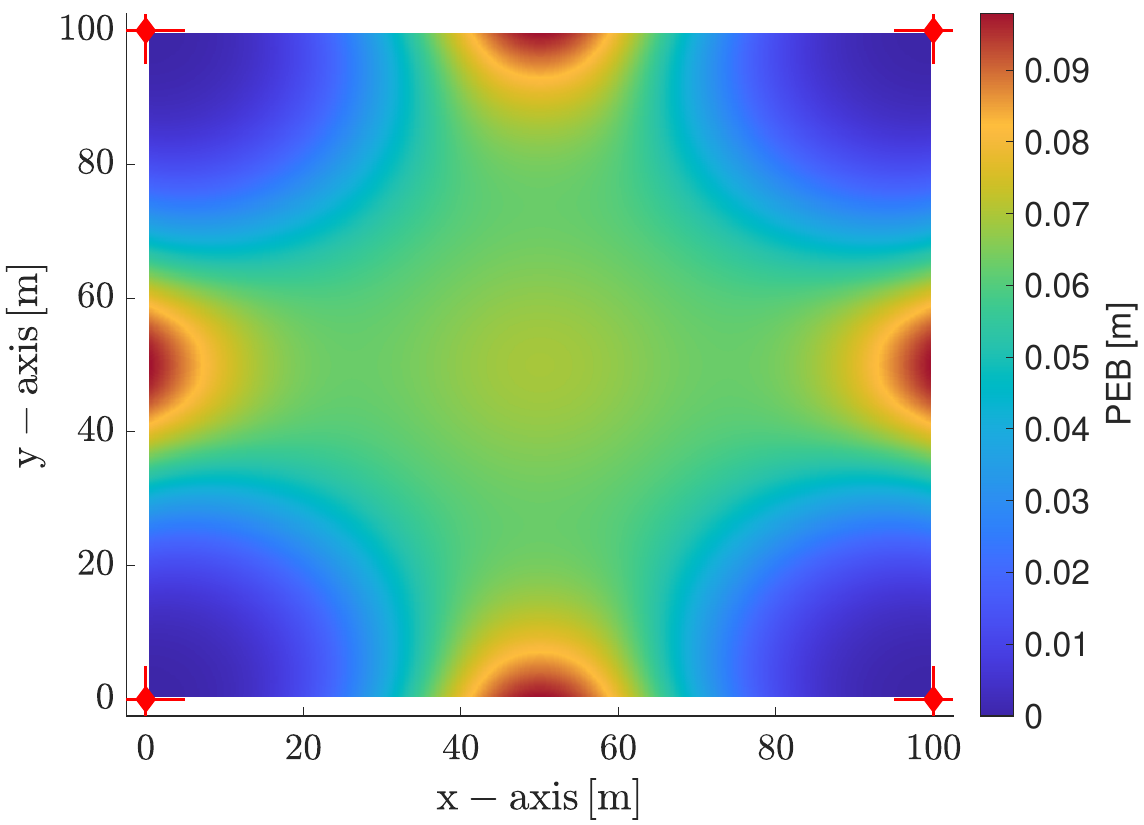}}\\ 
    \caption{Heatmaps showing the \ac{PEB} of position estimation for a \ac{JSC} network composed of $\Nbs = \{2, 3, 4\}$ monostatic \acp{BS}, in a), b), and c), respectively. The heatmaps are computed inside the monitored area according to the system parameters in Table~\ref{tab:sim_param}. The active \acp{BS} are indicated by red markers.}
    \label{fig:heatmamps}
\end{figure*}

\begin{figure*}[ht]
\captionsetup{font=footnotesize,labelfont=footnotesize}
    \centering
    \subfloat[][PEB vs $K$ \label{fig:PEBvsK}] 
    {\includegraphics[width=0.325\textwidth, keepaspectratio]{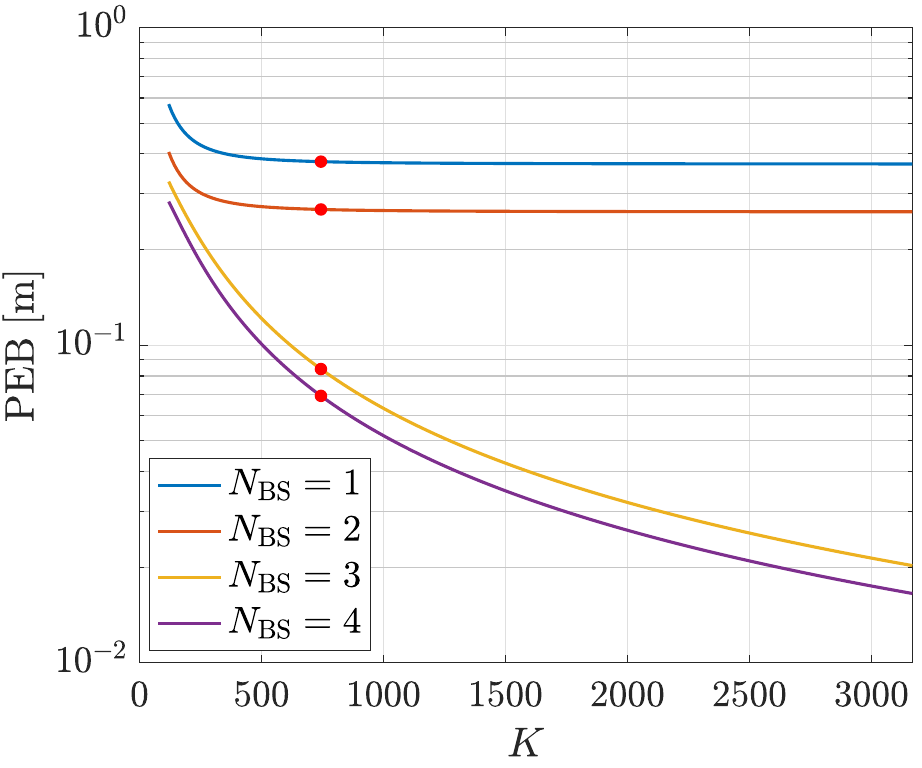}} 
    \,
    \subfloat[][PEB vs $\Nr$ \label{fig:PEBvsNr}]
    {\includegraphics[width=0.33\textwidth, keepaspectratio]{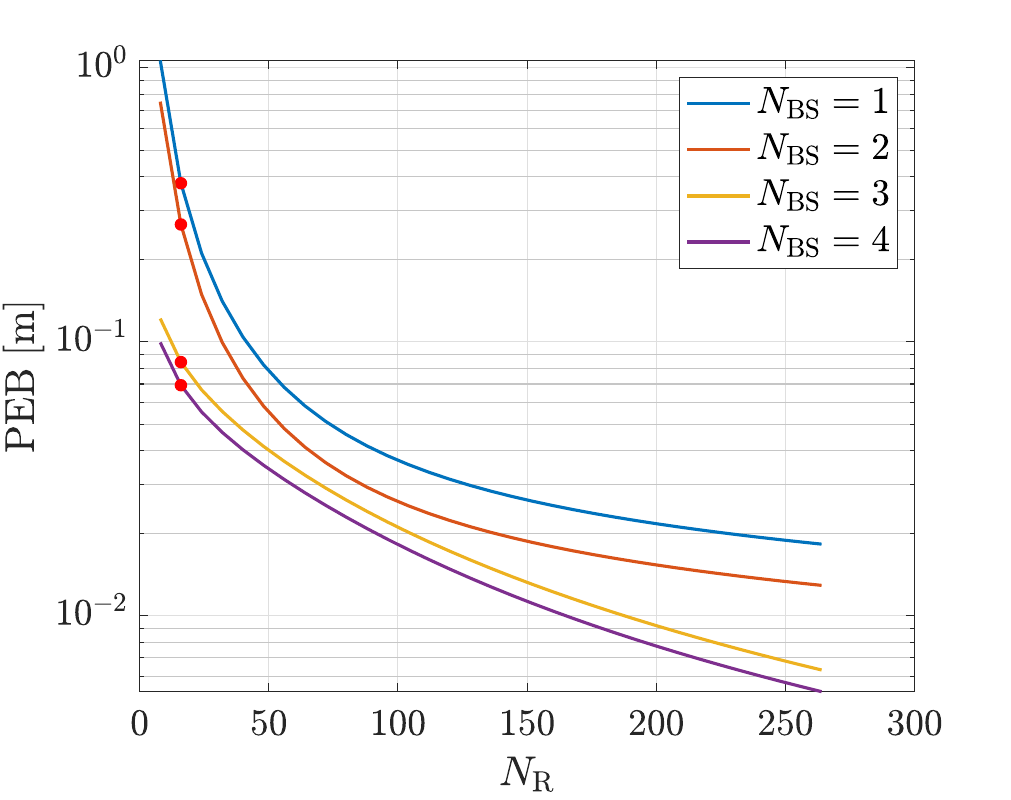}}
    \,
    \subfloat[][PEB vs $\rho$ \label{fig:PEBvsRho}]
    {\includegraphics[width=0.325\textwidth, keepaspectratio]{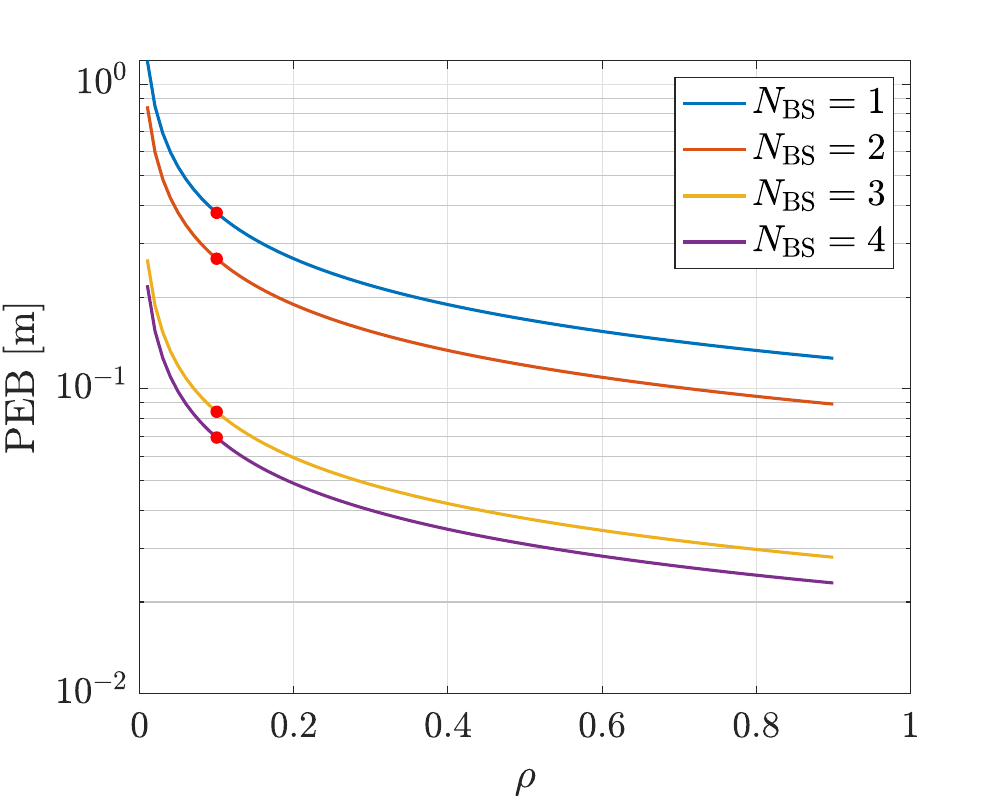}}\\ 
    \caption{PEB as a function of: a) the number of active subcarriers $K$, b) the number of antennas $\Nr$ at \ac{Rx}, c) the fraction of power reserved for sensing $\rho$. Red dots correspond to the PEB calculated by fixing the values of $K$, $\rho$, and $\Nr$ as in Table~\ref{tab:sim_param}. Also, the number of \acp{BS}, $N_\mathrm{BS}$, is varied.}
    \label{fig:PEBAnalysis}
\end{figure*}
\section{Numerical Results}\label{sec:numres}
In this section, we provide some numerical results to analyze the fundamental limits of target localization accuracy for a cooperative \ac{JSC} network. 
The analytical framework introduced in Section~\ref{sec:PEB} is used to evaluate the \ac{PEB} in the presence of a variable number $N_\mathrm{BS}$ of \acp{BS}, up to $4$, considering a single scatterer scenario and the system parameters in Table~\ref{tab:sim_param}. 
Without loss of generality, we consider the \acp{BS} to be positioned at the corners of a square area of side $100\,$m, with the \ac{Tx}/\ac{Rx} \acp{ULA} of each \ac{BS} pointing toward the center of the square, as shown in Fig.~\ref{fig:cooperative_scenario}. In particular, starting from the \ac{BS} in the lower left corner of the square and moving counterclockwise, the following vector of rotation angles is considered $\boldsymbol{\vartheta} = \{\pi/4, \pi-\pi/4, \pi + \pi/4, -\pi/4 \}$. 

Two types of analysis are conducted. The first involves varying the target position within the designated area while keeping all the system parameters constant. Conversely, the target position remains fixed in the second analysis while one parameter among $K$, $\rho$, and $\Nr$, is varied.
\begin{table} [t] 
\centering
 \caption{JSC network parameters} \label{tab:sim_param}
 \resizebox{0.8\columnwidth}{!}{
 \renewcommand{\arraystretch}{1}
 \begin{tabular}{l |l |l}
\toprule
Parameters [5G NR] & Symbols & Values \\
\midrule
Number of \ac{Rx} antennas & $\Nr$ & $16$ \\
Carrier frequency & $f_\mathrm{c}$ & $28$ GHz \\
OFDM symbols (time slots) & $M$ & $112$\\
Active subcarriers & $K$ & $744$ \\
Subcarrier spacing & $\Delta f$ & $120$ kHz \\
Total EIRP & $P_\mathrm{T} G^a_\mathrm{T}$ & $30$ dBm\\
Fraction of power for sensing & $\rho$ & $0.1$ \\
Power spectral density & $N_0$ & $4\cdot 10^{-20}$ W/Hz\\
Target radar cross-section & $\sigma$ & $1$ m$^2$\\
 \bottomrule
\end{tabular}
}
\label{NRparam}
\end{table}

\subsection{PEB vs target position}
This section analyzes target localization accuracy in the considered area using heatmaps. The target is moved within the area, and for each position, the \ac{PEB} is computed as described in Section~\ref{sec:netmonoCRLB}. It is important to note that for an unbiased estimator, the \ac{PEB} serves as a lower bound for the \ac{RMSE} of target position estimation. Thus, this analysis provides valuable insights into the coverage capabilities of a \ac{JSC} network with a specified number $N_\mathrm{BS}$ of monostatic \acp{BS}. For instance, by setting a predefined minimum \ac{PEB}, one can perform a coverage analysis of the sensing functionality.

The results are presented in Fig.~\ref{fig:heatmamps}, where the \ac{PEB} is computed for cooperative scenarios involving two, three, and four \acp{BS}. Notably, satisfactory localization accuracy can be achieved with just two \acp{BS}, attaining a maximum \ac{PEB} of approximately $0.25\,$m. This is observed even in the worst-case scenario, where the target is equidistant from both \acp{BS}, as illustrated in Fig.~\ref{fig:heatmap2}. Furthermore, the maximum \ac{PEB} decreases by approximately 30\% and 60\% when three and four \acp{BS} cooperate, as shown in Fig.~\ref{fig:heatmap3} and Fig.~\ref{fig:heatmap4}, respectively.

\subsection{PEB vs system parameters}
In this analysis, the target position is fixed at the center of the square, specifically at coordinates $(50, 50)\,$m, and the \ac{PEB} is computed while varying $K$, $\rho$, and $\Nr$. The aim is to explore which parameters prominently influence localization accuracy. The results are shown in Fig.~\ref{fig:PEBAnalysis} where red dots indicate \ac{PEB} values computed with all the parameters set according to Table~\ref{tab:sim_param}. Fig.~\ref{fig:PEBvsK} reveals that allocating many subcarriers for sensing may not be advantageous in cooperative networks with up to two \acp{BS}.
Conversely, increasing the number of \acp{BS} significantly improves localization accuracy. 
In Fig.~\ref{fig:PEBvsNr}, the impact of the number of antennas at the \ac{Rx}, $N_\mathrm{R}$, is shown by considering up to $264$ antenna elements. Notably, increasing $N_\mathrm{R}$ greatly enhances localization performance, enabling sub-centimeter accuracy when $\Nr \geq 180$ and $N_\mathrm{BS} \geq 3$. 

Lastly, Fig.~\ref{fig:PEBvsRho} shows the impact of the fraction of power reserved for sensing. It is possible to observe that increasing $\rho$ notably affects overall accuracy, although less than increasing $\Nr$. However, selecting high $\rho$ is discouraged due to its adverse effect on communication performance resulting from power allocation. As for the previous analysis, the benefit of cooperation is evident, as increasing $N_\mathrm{BS}$ leads to lower \ac{PEB}.

\section{Conclusion}\label{sec:conclu}
In this work, we have derived and explored the fundamental limits of target localization accuracy within a cooperative \ac{JSC} network comprising multiple monostatic \ac{MIMO}-\ac{OFDM} \acp{BS}. Utilizing the additivity property of Fisher information, we derived a framework to assess target positioning accuracy, using the \ac{PEB} metric, for an arbitrary geometrical configuration. Our numerical analyses underscore the advantages of cooperative approaches, revealing that an increase in the number of \acp{BS} enhances localization performance and the sensing system's coverage despite constraints on power, bandwidth, and antenna arrays. Notably, our results indicate that the number of receiving antennas is pivotal in boosting localization accuracy.


\balance
\bibliographystyle{./bibliography/IEEEtran}
\bibliography{./bibliography/IEEEabrv,./bibliography/Bibliography}

\end{document}